\documentclass[final,preprint,aps,showpacs,fleqn]{revtex4}
\usepackage{bm}
\usepackage{latexsym,amssymb,amsfonts,amsmath}
\usepackage{graphicx}
\usepackage{setspace}

\begin{document}

\title{Heavy quarkonium and dynamical gluon mass at
non-zero temperature in instanton vacuum model
}
\author{
M. Musakhanov\footnote{Presented at Diffraction and Low-x 2018
 26 Aug. 2018 -- 1 Sept. 2018 
Reggio Calabria, Italy},
Sh. Baratov, N. Rakhimov}
\affiliation{
Theoretical  Physics Dept, 
Uzbekistan National University,\\
 Tashkent 100174, Uzbekistan
}

\begin{abstract}
In the framework of the Instanton Liquid Model we evaluate the heavy
quark $\bar{Q}Q$ potential at nonzero temperature $T$. The potential
has two components:  contribution due to direct interaction with instantons,
 and the modification of the one-gluon exchange contribution via instanton-generated
dynamical \char`\"{}electric\char`\"{} gluon mass $M_{el}(q,T)$.
 We conclude that the nonperturbative ILM contributions to the $Q\bar{Q}$
potential have pronounced temperature dependence, which might be tested
in phenomenological analyses of charmonia production data.
\end{abstract}
 
\maketitle

\section{Motivation and Introduction}

Heavy quarkonium $Q\bar{Q}$ states created  in high energy  heavy
ion collisions can be used as a thermometer of the hot Quark-Gluon
Plasma formed at later stages due to final state interactions. For
this reason, the analysis of heavy quark dynamics occupies one of
the central places in hot matter studies~\cite{Schmidt:2018rkw}.  In the heavy quark mass
limit, the dynamics of $\bar{Q}Q$ pair is perturbative, however numerically
the mass of charmed quark $m_{c}\approx1.27\,{\rm GeV}$ is not very
large, and for this reason nonperturbative effects might be important,
though to a lesser degree than for the light quarks. For this reason
heavy quarks might be used as a probe sensitive to the onset of the
nonperturbative dynamics, and seek for one of the  constituents of
QCD vacuum, the  instantons. The distribution of instantons in the
QCD vacuum is described in the Instanton Liquid Model (ILM) framework.
At nonzero temperature $T\neq0$ this framework  predicts the temperature
dependence of the mean instanton size ${\bar{\rho}}(T)$ and density
$n(T)$, and thus provides an effective approach which might be considered
as a model of Quark-Gluon plasma.\\

In this proceeding we present results for the temperature-dependent
$Q\bar{Q}$ potential due to interactions with instantons. We found
that the ILM-induced potential differs significantly from the perturbative
Coulomb-like shape, has a pronounced temperature dependence and thus
affects the observed quarkonia yields.


\centerline{\textbf{\emph {Instanton Liquid Model}}}

\textbf{Instanton Liquid Model (ILM) at zero temperature $T=0.$}~
As was discussed in~\cite{Schafer:1996wv,Diakonov:2002fq}, the ILM
manages to describe all the nonperturbative physics using only two
main parameters   the average instanton size ${\bar{\rho}}$ and density
$n$.  These parameters have been estimated independently in different
approaches: the phenomenological~\cite{Schafer:1996wv} and variational~\cite{Diakonov:2002fq}
estimates of these parameters give $n^{-1/4}=R\approx1\,{\rm fm}$,
${\bar{\rho}}\approx0.33\,{\rm fm}$; the lattice studies~\cite{lattice}
result in $R\approx0.89\,fm$, $\bar{\rho}\approx0.36\,fm$, and the
 phenomenological estimates with account of $1/N_{c}$ corrections~\cite{Goeke:2007bj}
yield $R\approx0.76\,fm$, ${\bar{\rho}}\approx0.32\,fm$. As we can
see, all these estimates agree with each other within $10\%$ accuracy,
and in what follows for the sake of definiteness we will use for numerical
calculations the first set of parameters. The  packing fraction parameter
$\pi^{2}{\bar{\rho}}^{4}n\sim0.1$, which corresponds to the fraction
of the whole space occupied by instantons, is small, which justifies
 independent averaging over the collective coordinates of instantons.
Due to interactions with instanton ensemble, the  light quarks acquire
the dynamical  mass $M\sim({\rm packing\,parameter})^{1/2}{\bar{\rho}}^{-1}\approx365$~${\rm MeV}$.
The ILM approach manages to describe reasonably well a large number
of low-energy observables involving light hadrons~\cite{Goeke:2007bj}.

\textbf{ILM instanton size vs light hadron and heavy quarkonium $Q\bar{Q}$
sizes.} The average instanton size $\bar{\rho}$ determines average
distance at which the instanton effects are pronounces. As could be
seen from the above-given estimates, this size is comparable to the
average size of the dynamic quark inside nucleon  is $r_{N}\sim0.3-0.45\,{\rm fm}$~\cite{Weise:1985cd},
and to the size of the $\bar{Q}Q$ charmonia, as could be seen from
the Table~\ref{tab:SizesCharm} (see~\cite{quarkoniumsize} for
details). For this reason we expect that the lowest quarkonia should
be significantly affected by the contributions of instantons. The
lowest charmonia due to their small size should be insensitive to
the large-distance confinement, and for this reason  we expect that
ILM framework supplemented with perturbative gluons should be applicable
for their description.
\begin{table}[h]
\begin{tabular}{|c|c|c|c||c|c|c|c|c|}
\hline 
State  & $J/\psi$  & $\chi_{c}$  & $\psi'$  & $\Upsilon$  & $\chi_{b}$ & $\Upsilon'$ & $\chi_{b}'$ & $\Upsilon^{''}$ \tabularnewline
\hline 
mass {[}Gev{]}  & 3.07  & 3.53  & 3.68  & 9.46  & 9.99 & 10.02 & 10.26 & 10.36 \tabularnewline
\hline 
size $r$ {[}fm{]}  & 0.25  & 0.36  & 0.45  & 0.14  & 0.22 & 0.28 & 0.34 & 0.39 \tabularnewline
\hline 
\end{tabular}{\caption{{\label{tab:SizesCharm}Sizes of charmonia evaluated
in potential models. Data are from~\cite{quarkoniumsize}.}}
}
\end{table}

\textbf{ILM at nonzero temperature.} The temperature dependence is
introduced in the model as (anti)periodical boundary condition for
all fields, with period in time $\beta=1/T$, and with additional
contributions due to fluctuations of thermal gluon (and light quarks). 
In the Figure~\ref{rho,n} below we demonstrate the temperature
dependence of the ILM parameters. We can see that the instanton gas
density and average instanton size decrease as a function of temperature
$T.$
\begin{figure}[h]
\centerline{\includegraphics[scale=0.43]{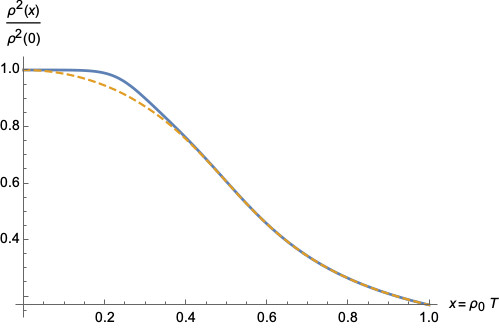}\includegraphics[scale=0.43]{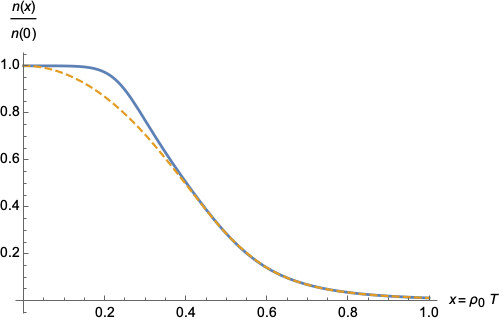}}
\label{rho,n} \caption{\textbf{\emph{Left}}:  The temperature dependence of the average instantons
size, normalized to unity at $T=0$ (${\bar{\rho}}^{2}(x={\bar{\rho}}T)/{\bar{\rho}(0)}^{2}$).
\textbf{\emph{Right:}} The temperature dependence of the instanton
density normalized to unity at $T=0$ ($n(x)/n(0)$).  In both plots
the dashed lines correspond to the full suppression of instantons
due to Debye screening~\cite{Diakonov:1988my}, whereas the full
line corresponds to  an interpolation between no suppression below
critical temperature $T_{c}\approx150\,{\rm MeV}$ and full suppression
above $T_{c}$, with a width $T=0.3\,T_{c}$~\cite{Schafer:1996wv}.
}
\end{figure}

\section{Gluons in ILM }

\begin{figure}[h]
{\includegraphics[scale=0.4]{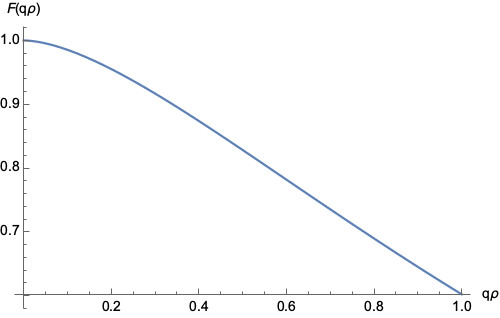}\includegraphics[scale=0.6]{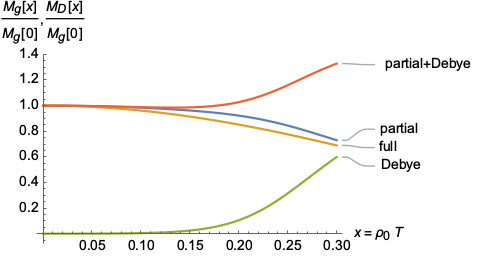}}
\caption{\label{fig:Mg}\textbf{\emph{Left:}} momentum $q$ dependence of the
gluon electric mass at $T=0$ normalized to unity at
$q=0$: $M_{el}(q,\,0)=M_{g}F(q).$ \textbf{\emph{Right:}} Temperature
dependence of the gluon effective mass $M_{g}$ and Debye screening
$M_{D}$, normalized to unity at $T=0$. We use notations $x=T{\bar{\rho}}$;
the critical temperature corresponds to $x_{c}=0.25$. For $x>x_{c}$
the contribution of thermal gluon (and quark) fluctuations to the
gluon propagator lead to so-called gluon Debye screening mass $M_{D}(x),$
and we use for it the lattice  parametrization $M_{D}(x)=1.52\,x/{\bar{\rho}}\Theta(x-x_{c})$~\cite{Silva:2013maa}.
The other notations are the same as in  the Fig.~\ref{rho,n}.}
\end{figure}

\textbf{$T=0$ case}~\cite{Musakhanov:2017erp}. The interaction
of gluons with instantons leads to generation of the dynamical momentum-dependent
gluon mass. In order to find it, we have to solve the zero-mode problem,
 to average the total gluon propagator in ILM using the method of.~\cite{DPP1989}
 extended to gluon sector, and finally to find gauge invariant dynamical
gluon mass $M_{g}(q)$ as a function of Euclidean momentum $q$. It
was found that the dynamical mass $M_{g}$ has a form $M_{g}(q)=M_{g}F(q),$
where the form-factor $F(q)=q{\bar{\rho}}K_{1}(q{\bar{\rho}})$ is
shown in the right panel of the Figure~\ref{fig:Mg}, and $M_{g}\sim M\sim({\rm packing\,parameter})^{1/2}{\bar{\rho}}^{-1}\approx362\,{\rm MeV}.$
The value of $M_{g}$ also determines the strength of gluon-instanton
interaction.

\textbf{ Nonzero temperature $T\neq0$}~\cite{Musakhanov:2018gho}.
In this case again we have to impose the time periodicity condition
for the fields, which breaks relativistic co variance of the gluon
propagator, as well as take into account the temperature induced modifications
of ILM parameters. For our studies the most important observable is
the temperature dependence of the so-called \char`\"{}electric\char`\"{}
dynamical gluon mass $M_{el}(q,T)$  shown in the right panel of the
Figure~\ref{fig:Mg}. We can see that this dependence is mild in
the region $T<T_{c}$.

\section{Singlet $Q\bar{Q}$ potential in ILM at $T\protect\neq0$}

\textbf{Direct instanton contribution to the singlet $Q\bar{Q}$ 
potential  at $T\neq0$}.
\begin{figure}[h]
\centerline{\includegraphics[scale=0.45]{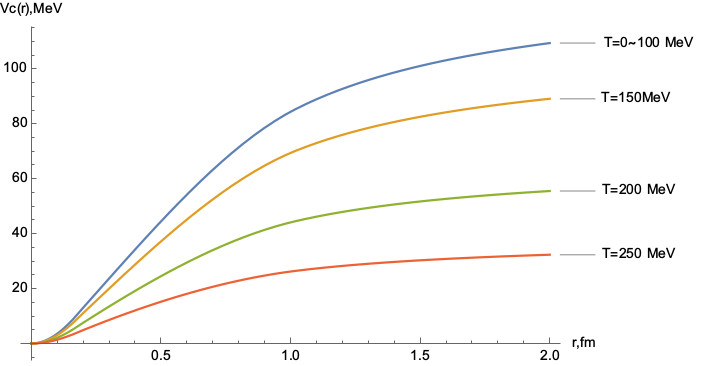}}
\caption{\label{fig:ColorSIngletV}The color singlet $\bar{QQ}$ potential
$V_{c}(r)$ at different temperatures $T$ evaluated in the instanton
liquid model. }
\end{figure}\vskip -0.3cm
In the Figure~\ref{fig:ColorSIngletV} we present out results for
the potential $V_{c}(r,T)$ evaluated from Wilson loop  in ILM, using
the method introduced in~\cite{DPP1989} . At zero temperature, the
potential has a finite limit,  $V_{c}(r\rightarrow\infty,T=0)=2\,\Delta m_{Q}\approx140\,{\rm MeV},$
 where $\Delta m_{Q}$ is the ILM contribution to the heavy quark
mass, and is due to the heavy quark-instanton interaction. 
As could be seen from the Table~\ref{tab:ILM-mass}, the direct ILM
effects are not small, are of order $\sim30\,\%$ in comparison with
the experimental data and strongly depend on the choice of ILM parameters.
\begin{table}[h]
\begin{center}
\scalebox{0.9}{
\begin{tabular}{c|c|c|c}
$\Delta M_{c\bar{c}}(J^{P})$ & Set~I & Set~II  & Exp.\tabularnewline
\hline 
$\Delta M_{\eta_{c}}(0^{-})$  & 118,81  & 203,64  & $433,6\pm0.6$\tabularnewline
$\Delta M_{J/\psi}(1^{-})$  & 119,57  & 205,36  & $546,916\pm0.11$\tabularnewline
$\Delta M_{\chi_{c0}}(0^{+})$  & 142,43  & 250,86  & $864,75\pm0.31$\tabularnewline
\end{tabular} 
}
\end{center}\vskip -0.4cm
\caption{\textbf{\label{tab:ILM-mass}}ILM direct instanton contribution to
the charmonium states at $T=0$~\cite{Turimov:2016adx}. $\Delta M_{c\bar{c}}=M_{c\bar{c}}-2m_{c}$
in {[}MeV{]}. $m_{c}=1275\,MeV.$ Set~I corresponds to the instanton
vacuum parameters ${\bar{\rho}}=0.33$~fm and ${R}=1\,\mathrm{fm},$
set~II uses values ${\bar{\rho}}=0.36$~fm, ${R}=0.89\,\mathrm{fm}.$}
\end{table}
\vskip -0.3cm

\begin{figure}[h]
\includegraphics[scale=0.55]{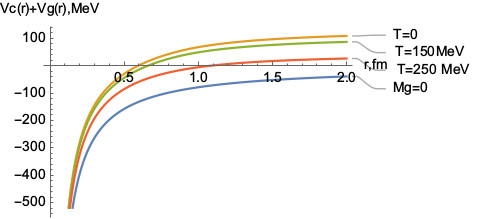}\includegraphics[scale=0.5]{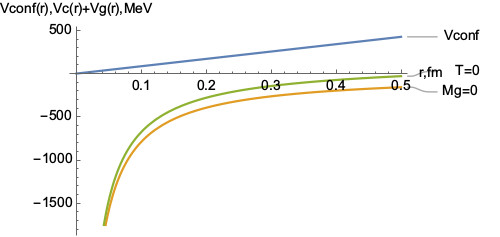}\caption{\label{fig:Comparison}~Left: Comparison of the full ILM potential
at different temperatures $T$ with Coulomb piece of Cornell potential  ($M_g=0$ line).
Right: comparison of ILM potential at zero temperatures with Coulomb 
and confinement pieces of the Cornell potential.}
\end{figure}

At very short distances, the dynamics of $\bar{Q}Q$ is described
by the pQCD, so we add to ILM potential $V_{c}(r)$ the one-gluon
exchange contribution 
\\
\centerline{$
V_{g}(r,T)=\lambda\cdot\bar{\lambda}\,g^{2}\int\frac{d^{3}k}{(2\pi)^{3}}\exp(i\vec{k}\vec{r})\left(\vec{k}^{2}+M_{el}^{2}(\vec{k},T)\right)^{-1},
$}
\\
where $\lambda\cdot\bar{\lambda}=-4/3$ and $+1/6$ for $Q\bar{Q}$
in color singlet and color octet states respectively and we use
gluon effective mass $M_{g}(q,\,T)$ evaluated earlier. 
Finally, in the Figure~\ref{fig:Comparison} we compare the full
ILM potential with the Cornell potential, which consists of a sum
of Coulomb-like and linearly growing confinement contributions~\cite{Bali:2000gf}.
We can observe that the potentials differ at large distances, which
implies that the $\bar{Q}Q$ phenomenology should be reanalyzed.

\section{Discussion}

The dynamics of the heavy$Q\bar{Q}$ pair is significantly affected
by the nonperturbative  QCD vacuum properties. We used the ILM as
the model of QCD vacuum,  both at zero temperature and in hot matter.
At $T=0$, heavy quark-instanton interaction generates nonperturbative
correction to the heavy quark mass, $\Delta m_{Q}\sim70\,{\rm MeV},$
whereas in case of the light quark-instanton and gluon-instanton interactions
the same interaction is more pronounced and gives the respective dynamical
masses $M\sim M_{g}\sim360\,{\rm MeV}.$ The instanton-induced nonperturbative
effects homogeneously decrease as a function of temperature $T$.

We also evaluated explicitly the  $Q\bar{Q}$ potential, which receives
sizable modifications due to direct interactions with instantons,
as well as generation of the  dynamical gluon mass in one-gluon exchange
contribution. The evaluated ILM potential differs significantly from
the phenomenological parametrizations available from the literature.
At zero temperature, our findings might be tested through studies
of the charmonia spectroscopy. At nonzero temperatures ($T\neq0$)
the average instanton size ${\bar{\rho}}(T)$ and density $n(T)$
are gradually decreasing functions of temperature, which leads to
pronounced temperature dependence of the $Q\bar{Q}$ potential. The
predicted $T$-dependence might be tested  heavy quarks production
processes in hadron-hadron collisions at high energies. 

\textbf{Acknowledgement.} M.M.  is thankful to Boris Kopeliovich and Marat Siddikov for the useful
 and helpful communications. This work is supported by Uz grant
 OT-F2-10. 

\end{document}